\documentclass[10pt,letterpaper]{article}
\usepackage{opex3}
\usepackage{cite}

\begin{document}

\title{Generation of cylindrical vector vortex beams by two cascaded metasurfaces}

\author{Xunong Yi,$^{1,3}$ Xiaohui Ling,$^{1}$ Zhiyou Zhang,$^{1}$ Ying Li,$^{1}$
Xinxing Zhou,$^{2}$ Yachao Liu,$^{2}$ Shizhen Chen,$^{2}$  Hailu
Luo,$^{1,2\ast}$ and Shuangchun Wen$^{1,2}$}

\address{$^1$ Shenzhen University-National University of Singapore Collaborative
 Innovation Center for Optoelectronic Science and Technology, Shenzhen University, Shenzhen 518060, China\\
$^2$ Laboratory for Spin Photonics, College of Physics and
Microelectronic
Science, Hunan University, Changsha 410082, China\\
$^3$ College of Physics and Electronic Information Engineering,
Hubei Engineering University, Xiaogan 432000, China}

\email{$^\ast$hailuluo@hnu.edu.cn} 


\begin{abstract}
We present a simple and efficient method to generate any cylindrical
vector vortex (CVV) beams based on two cascaded metasurfaces. The
metasurface works as a space-variant Panchratnam-Berry phase element
and can produce any desirable vortex phase and vector polarization.
The first metasurface is used to switch the sign of topological
charges associated with vortex, and the second metasurface is
applied to manipulate the local polarization. This method allows us
to simultaneously manipulate polarization and phase of the CVV
beams.
\end{abstract}



\section{Introduction}
Polarization and phase are two fundamental features of
electromagnetic waves. Conventional polarization states, such as
linear, circular, and elliptical polarizations, are spatially
homogeneous. Recently, light beams with a spatial inhomogeneous
polarization distribution (vector beams) have drawn much
attention~\cite{Zhan2009}. Most existing researches were devoted to
the case of polarization state with cylindrical symmetry
distribution (cylindrical vector beam) because of their distinctive
properties and potential
applications~\cite{Youngworth2000,Zhan2002,Dorn2003,Zhan2004,Kawauchi2007,Deng2007,Kozawa2010,Ye2014}.
Comparing with the conventional homogeneous polarization represented
by the fundamental Poincar\'{e} sphere and orbital Poincar\'{e}
sphere~\cite{Padgett1999,Galvez2003,Karimi2010}, the cylindrical
vector beams can be represented by a higher order Poincar\'{e}
sphere~\cite{Holleczek2011,Milione2011,Liu2014}. On the other hand,
optical vortices are intriguing optical structures with spiral
wavefronts. The phase around the vortex is characterized by
$\exp(im\varphi)$, where $m$ is the topological charge and $\varphi$
the azimuthal angle. Such beams with optical vortex phase carry
orbital angular momentum (OAM) of $m\hbar$ per
photon~\cite{Allen1992}. Vortex beams have found wide applications
in optical manipulation, imaging, and optical
communication~\cite{Yao2011}. If an optical beam possesses both
vector polarization and helical phase, referred to as cylindrical
vector vortex (CVV) beams, it can provide more degrees of freedom
for beam manipulation~\cite{Zhao2013}.

Driven by their fantastic features and various applications, several
methods have been proposed to produce CVV beams, such as a
liquid-crystal polarization converter~\cite{Cardano2012},
interference of different vector modes~\cite{Viswanathan2009},
spatial light modulator~\cite{Chen2011}, spiral phase
plates~\cite{Niv2006}, and forked grating~\cite{Galvez2012}.
Recently, metasurface, a two-dimensional electromagnetic
nanostructure, is expected to expand the capabilities of existing
vortex optics for simultaneous control of intensity, polarization,
and phase~\cite{Litchinitser2012,Kildishev2013}. A compact
metasurface which consists of two concentric rings in a gold film
has been proposed to generate the OAM-carrying vector
beams~\cite{Zhao2013}. In addition, it has been demonstrated that
the polarization states can be controlled by a metasurface
fabricated by femtosecond laser writing of self-assembled
nanostructures in silica glass~\cite{Beresna2011}. However, the
vortex beam in this situation is confined to the two special cases:
the radial and the azimuthal polarizations.

In this work, we present a technique to generate CVV beams with a
simple optical device based on two cascading metasurfaces. The
metasurface is structured by writing a space-variant nanograting in
a silica glass. The nanograting structure results in a space-variant
effective birefringence. By tunably controlling the local
orientation and geometrical parameter of the nanograting, one can
achieve any desired polarization
manipulation~\cite{Bomzon2002,Levy2004,Lerman2011}. The two
dimensional nanostructures have a high transmissivity that allows us
to cascade the metasurfaces. The experimental equipment consisted of
two metasurfaces can generate CVV beams with arbitrary polarization
orientation and allows us to switch the signs of topological charges
of vortex by controlling the handedness of incident circular
polarization.

\section{Metasurface and Pancharatnam-Berry phase }

\begin{figure}
\centerline{\includegraphics[width=12cm]{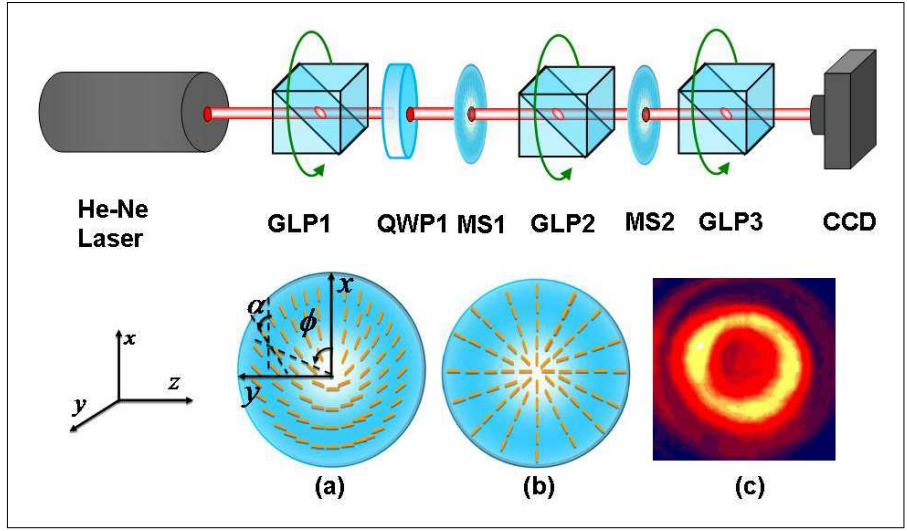}}
\caption{\label{Fig1} Experimental setup for generating arbitrary
CVV beams. Inset (a) and (b): Schematic pictures of metasurfaces
with $q=0.5$ and $q=1$, respectively. Space-varying nanograting
consisting of nanoscale waveplate with a spatially varying fast axis
directions. Inset (c): Intensity distribution of the CVV beams
presents doughnut profiles.}
\end{figure}

A metasurface is fabricated by femtosecond laser writing of
self-assembled nanostructures in silica glass~\cite{Beresna2011}.
The writing pattern is locally varying and of the order less than a
wavelength. This leads to build an artificial uniaxial crystal with
homogeneous phase retardation $\delta$ at the working wavelength
$\lambda$ and locally varying optical axes direction parallel and
perpendicular to the writing direction of the nanostructure in the
transverse plane ($xy$). As the dimension of the nanograting
structure is smaller than the incident wavelength, only the zero
order is a propagating order, and all other orders are
evanescent~\cite{Niv2008}.

In particular, the optical axis direction can be specified by the
following expression:
\begin{equation}
\alpha(r,\varphi)=q \varphi +\alpha_0,
\end{equation}
where ($r$, $\varphi$) is the polar coordinate representation,
$\alpha_0$ is a constant angle specifying the initial orientation
for $\varphi=0$, and $q$ is a constant specifying the topological
charge of the metasurface. The manipulation of polarization state
and wavefront is obtained by using the effective birefringence of
the nanostructure.

\begin{figure}
\centerline{\includegraphics[width=11cm]{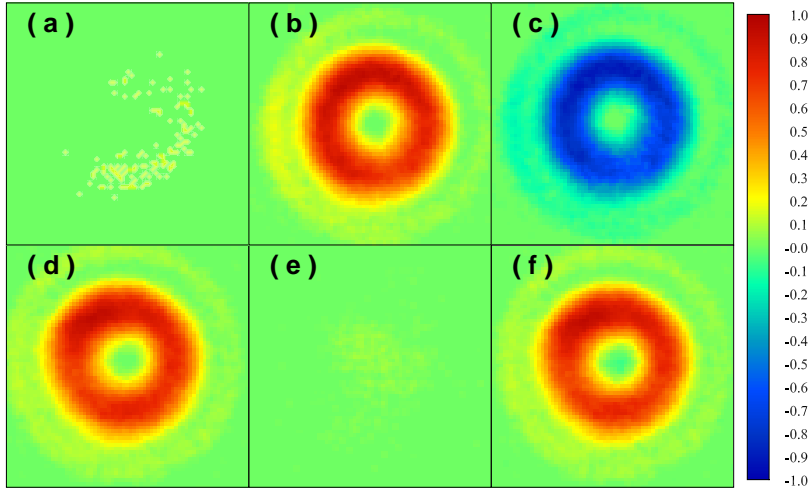}}
\caption{\label{Fig2} Stokes parameter $S_3$ of the beams emerging
from MS1. Row 1: the experimental results when QWP1 is placed at
$45^\circ$. (a) and (b) are the intensity distributions when GLP4 is
placed at $\pm 45^\circ$, respectively. (c) the corresponding Stokes
parameter $S_3$. Row 2: the experimental results when QWP1 is placed
at $-45^\circ$. (d) and (e) are the intensity distributions when
GLP4 is placed at $\pm 45^\circ$, respectively. (f) the
corresponding Stokes parameter $S_3$.}
\end{figure}

If the orientation of the nanotructure is space-variant at each
location, the grating can be described by the coordinate-dependent
matrix~\cite{Marrucci2006,Ling2012,Marrucci2011}:
\begin{equation}
\mathbf{T}(r,\varphi)=\mathbf{M}(r,\varphi)\mathbf{J}\mathbf{M}^{-1}(r,\varphi).
\end{equation}
Here, $\mathbf{J}$ is the Jones matrix of a uniaxial crystal, and
\begin{equation}
\mathbf{M}(r,\varphi)=\left(
\begin{array}{cc}
\cos\alpha & \sin\alpha \\
\sin\alpha & -\cos\alpha
\end{array}
\right)\label{Jones},
\end{equation}
where $\alpha(r, \varphi)$ is the local optical axis orientation of
the metasurface. It is easily derived that the Jones matrix
$\mathbf{T}(r,\varphi)$ can be written as
\begin{equation}
\mathbf{T}(r,\varphi)=\cos\frac{\delta}{2}\left(
\begin{array}{cc}
1 & 0 \\
0 & 1
\end{array}
\right)+i\sin\frac{\delta}{2}\left(
\begin{array}{cc}
\cos2\alpha & \sin2\alpha \\
\sin2\alpha & -\cos2\alpha
\end{array}
\right)\label{Jones}.
\end{equation}

Let us now consider that the metasurface is normally illuminated by
a circularly polarized vortex wave with spin angular momentum (SAM)
$\sigma\hbar$ and OAM $m\hbar$, where $\sigma=+1$ for the
left-handed circular (LHC) case and $\sigma=-1$ for the right-handed
circular (RHC) one. Its Jones vector is then given by
$\mathbf{E}_{in}(r,\varphi)=E_0(r,\varphi)\times[1, \sigma
i]\exp(im\varphi)$. The output beam
$\mathbf{E}_{out}(r,\varphi)=\mathrm{T}(r,\varphi)\mathbf{E}_{in}(r,\varphi)$
can be written as
\begin{equation}
\mathbf{E}_{out}(r,\varphi)=E_0\cos\frac{\delta}{2}\exp(im\varphi)\left(
\begin{array}{cc}
1 \\ \sigma i
\end{array}
\right)+iE_0\sin\frac{\delta}{2}\exp[i(m\varphi+2\sigma\alpha)]\left(
\begin{array}{cc}
1 \\ -\sigma i
\end{array}
\right)\label{CVVTI}.
\end{equation}
We find that the output field can be regarded as a superposition of
a part that has the same SAM and OAM as the input one, and another
part having a reversed SAM and a modified OAM given by
$(m+2{\sigma}q)\hbar$. It means that the input light beam only
partially converts its spin angular momentum to its orbital
part~\cite{Marrucci2011}. When $m=0$, Eq.~(\ref{CVVTI}) represents
the case of that the metasurface is illuminated by a circularly
polarized light. The amplitudes of these two components of the
output field depend on the birefringent retardation $\delta$, i.e.,
$\cos\delta/2$ and $\sin\delta/2$, respectively.

More interestingly, the output field acquires a nonuniform phase
retardation. In general, any desirable wavefront can be generated by
designing the metasurface geometry. The second term of
Eq.~(\ref{CVVTI}) indicates a polarization orthogonal to that of the
input wave, and its phase is twice of the local optical axis
orientation $2\sigma\alpha $. Similarly, the wavefront with an
opposite topological charge should be generated if the input
polarization handedness is inverted. This particular approach for
optical wavefront shaping is purely geometric in nature, and the
additional phase factor is the so-called Pancharatnam-Berry
geometric phase~\cite{Marrucci2006}. The wavefront-shaping device
based on this principle have been realized and can be referred to as
Pancharatnam-Berry phase optical element~\cite{Bomzon2002}.

\begin{figure}
\centerline{\includegraphics[width=12cm]{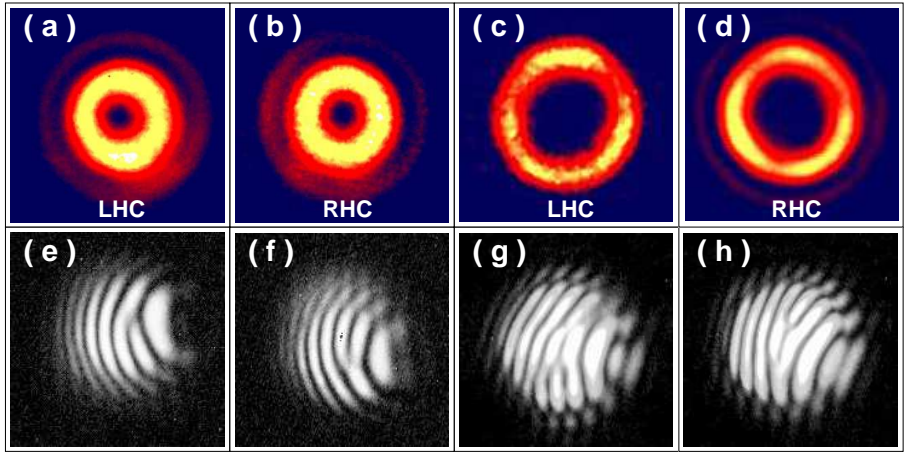}}
\caption{\label{Fig3} The intensity distribution (upper panels) of
the beam emerging from the first metasurface and the interference
patterns (lower panels) after their superposition with the reference
beam. (a) and (b) are intensities for the LHC and RHC polarization
incidence, and (e) and (f) are interference patterns with the
reference beam, respectively, for $m=1$. (c) and (d) as well as (g)
and (h) are the corresponding results for $m=2$.}
\end{figure}

As a special case, a homogeneous half waveplate can completely
convert a LHC (RHC) polarization into a RHC (LHC) polarization. The
circular polarized input light reverses its handedness and acquires
a constant phase factor when passing through the half-wave plate.
But the inhomogeneous anisotropic metasurface can result in the
occurrence of the spin-orbit interaction~\cite{Niv2008}. As
expected, a half-wave metasurface ($\delta=\pi$) not only can invert
the handedness of circular polarization, but also applies the output
wave an azimuthal phase, i.e., OAM. It is well known that LHC and
RHC polarized light possess angular momentum of $\pm\hbar$ per
photon. From Eq.~(\ref{CVVTI}), a LHC polarized incident light can
be transformed into a RHC polarized vortex light. Thus the
spin-orbit interaction in the metasurface leads to the appearance of
the helical phase with topological charge $2q$. Similarly, for the
right-handed circular polarization incidence, a left handed circular
polarization vortex light with topological charge $-2q$ will be
gotten.

Let us now consider that the metasurface is illuminated by a linear
polarized vortex wave whose Jones vector is then given by
$\mathbf{E}_{in}(r,\varphi)=E_0(r,\varphi)\times[\cos\theta,
\sin\theta ]\exp(im\varphi)$. The output beam
$\mathbf{E}_{out}(r,\varphi)=\mathrm{T}(r,\varphi)\mathbf{E}_{in}(r,\varphi)$
can be obtained as
\begin{equation}
\mathbf{E}_{out}(r,\varphi)=E_0(r,\varphi) e^{im\varphi}\left(
\begin{array}{cc}
\cos(2\alpha-\theta) \\
\sin(2\alpha-\theta)
\end{array}
\right)\label{CVVTII}.
\end{equation}
Here, we have assumed $\delta=\pi$. From Eq.~(\ref{CVVTII}) we can
see that the output beam is a CVV beam. It means that the
metasurface can apply to change the spatial distribution of
polarization. This interesting phenomenon can be explained as the
appearance of the Pancharatnam-Berry phase when the beam passes
through the metasurface~\cite{Marrucci2006,Marrucci2011}. By
rotating the incident polarization, the polarization direction of
the CVV beam can be tunable effectively.

\section{Experimental results and discussions}

Figure~\ref{Fig1} shows the schematic illustration of the
experimental setup for generating arbitrary CVV beams. In the
experiment, we use a paraxial Gaussian beam as the light source,
which is produced by a He-Ne laser working at the wavelength
$\lambda=632.8~nm$. The laser beam is converted to linear
polarization state by a Glan laser polarizer (GLP1). The angle
between the fast axis of the quarter-wave plate (QWP1) and the
transmission axis of the GLP1 is arranged at $45^\circ$. The linear
polarized beam is converted into a left-handed circular polarization
by QWP1. The first metasurface (MS1) is used to transform the LHC
(RHC) polarization beam into RHC (LHC) polarization with a helical
phase. Figure~\ref{Fig1}(a) and~\ref{Fig1}(b) present two schematic
pictures of the metasurface. In our scheme, the metasurface is
fabricated by femtosecond laser writing of self-assembled
nanostructures in a silica glass (Altechna). In principle, the
metasurface can be regarded as a two dimensional aperiodic
nanograting, which can produce space-variant phase and polarization.

As mentioned above, the QWP1 in our experimental setup is placed at
$\pm 45^\circ$ in order to generate a left- or right-handed
circularly polarized beam. And then, the first metasurface will
convert the LHC (RHC) beam into a RHC (LHC) polarized vortex beam.
To obtain the polarization distribution, the Stokes parameter are
measured~\cite{Born1997}. As shown in Fig.~\ref{Fig1}, we can add
another Glan laser polarizer (GLP4) and another quarter waveplate
(QWP2) behind the MS1. By rotating the GLP4 to two angles
($\pm45^\circ$) and holding QWP2 in the $x$ direction, we can obtain
the intensity distributions on the CCD. As the Stokes parameter
$S_3$ is defined as
$S_3=(I_{+45^\circ}-I_{-45^\circ})/(I_{+45^\circ}+I_{-45^\circ})$,
where $I_{\pm45^\circ}$ represents the recorded intensity when the
transmission axis of the GLP4 is set as $\pm45^\circ$. After a
series of data process, we can obtain the $S_3$ pixel by pixel. The
experimental results indicate that the MS1 transforms the LHC
polarization beam into the RHC polarization one and vice versa, as
shown in Fig.~\ref{Fig2}.

\begin{figure}
\centerline{\includegraphics[width=9cm]{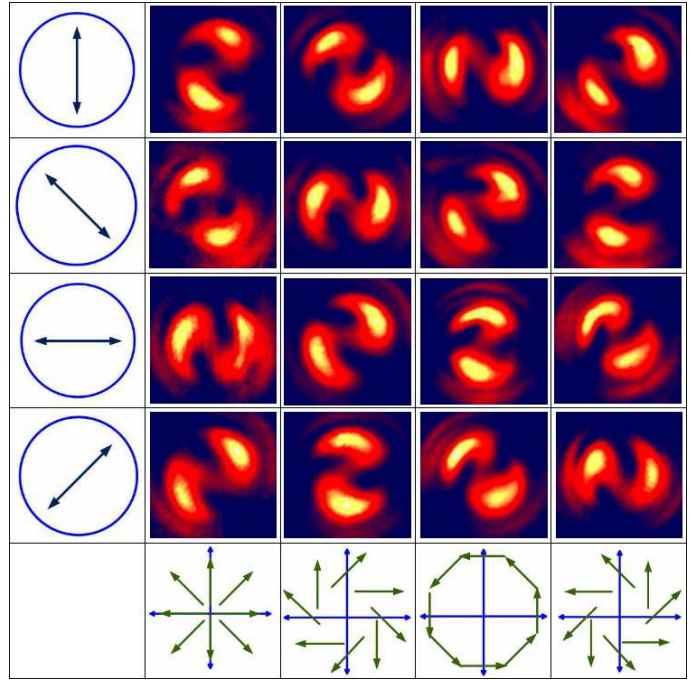}}
\caption{\label{Fig4} A set of experimentally generated CVV beams
when GLP1 is placed at $+45^\circ$. The first column shows the
direction of the linear polarizer (GLP3). The next four columns show
the intensity distributions behind the polarization analyzer at
different polarization angles. The lowest row sketches the
reconstructed vector fields of the output beams}
\end{figure}

To observe the phase structure of the generated vortex beams, we
measure the interference pattern between the vortex beam and the
reference beam which is split from the input Gaussian beam. The
second metasurface is applied to change the spatial distribution of
polarization. By rotating GLP2, the polarization direction of the
CVV beam can be easily modulated. The polarizer (GLP3) acts as a
polarization analyzer, and the CCD camera is used to record the
transmitted intensity distribution. The experimental results are
plotted in Fig.~\ref{Fig3}. Figure~\ref{Fig3}(a) and~\ref{Fig3}(b)
show the intensity distribution of the RHC and the LHC polarized
vortex beam with topological charge $m=1$, respectively. The
interference patterns of the LHC (RHC) polarized incident case with
the reference beams are shown in Fig.~\ref{Fig3}(e)
and~\ref{Fig3}(f). This fork-like patterns indicate the phase
vortexes in the LHC (RHC) polarized beams. Additionally, we also
implement a group of experiments for $m=2$.

The second metasurface (MS2) is applied to change the spatial
distribution of polarization. We set the transmission axis of the
GLP2 inclined an angle of $\theta$ respect to the $x$ direction. The
emerging beam from the GLP2 can be written as $E_0(r,\varphi)
e^{i\varphi} [\cos\theta, \sin\theta]$. This is a linear
polarization vortex beam. From Eq.~(\ref{CVVTII}), the expression of
the output beam after the MS2 can be written as
\begin{equation}
\mathbf{E}_{out}^{II}(r,\varphi)=E_0(r,\varphi) e^{i\varphi}\left(
\begin{array}{cc}
\cos(2\varphi-\theta) \\
\sin(2\varphi-\theta)
\end{array}
\right)\label{CVVEII}.
\end{equation}
From Eq.~(\ref{CVVEII}), we can see that the output field is a CVV
beam with topological charge $+1$. One can obtain a vortex beam with
arbitrary polarization direction by rotating the GLP2. In principle,
the incident linear polarized beam can be regarded as a
superposition of LHC polarized and RHC polarized waves. Therefore,
the output CVV beam is the coherent superposition of a LHC polarized
vortex and a RHC polarized vortex beam [Eq.~(\ref{CVVTI})]. That is
to say, there appears a spin-dependent phase, the so-called
Pancharatnam-Berry phase, when the beam passes through the MS2.

\begin{figure}
\centerline{\includegraphics[width=9cm]{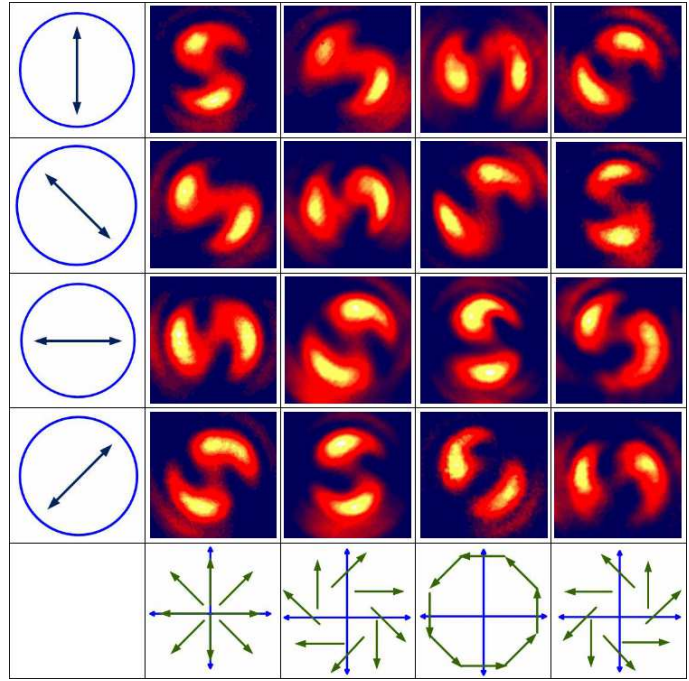}}
\caption{\label{Fig5} A set of experimentally generated CVV beams
when GLP1 is placed at $-45^\circ$. The first column shows the
direction of the linear polarizer (GLP3). The next four columns show
the intensity distributions behind the polarization analyzer GLP3 at
different polarization angles.}
\end{figure}

We now consider that a RHC polarized beam illuminates the MS1 by
setting the transmission axis of the GLP1 at $+45^\circ$.
Figure~\ref{Fig4} shows the measured intensity distributions when
the GLP2 is rotated at different angles. According to the measured
intensity distributions by rotating the polarization analyzer GLP3,
the polarization states of the output beams are sketched in the
lowest row. In Fig.~\ref{Fig4}, we can clearly observe a typical
anti-$``\emph{s}"$-shape pattern which is led by the helical phase
with topological charge $-1$. The mode in the second column
corresponds to a radially polarized vortex beam with GLP2 placed at
$0^\circ$ angle. Rotating the GLP2 at $45^\circ$, we can obtain a
polarized vortex beam with local polarization direction rotated
$45^\circ$ as shown in the third column. When GLP2 is placed at
$90^\circ$, a vortex beam with azimuthal polarization can be
obtained as displayed in the fourth column. And the fifth column
corresponds to a polarized vortex beam with GLP2 placed at
$-45^\circ$.

We also carry out another group of experiment when GLP1 is placed at
$-45^\circ$. This means that a LHC polarized beam illuminates on the
MS1. The experimental results are shown in Fig.~\ref{Fig5}. We find
that the intensity pattern is changed into $``\emph{s}"$ shape. It
suggests that the topological charge of the helical phase is $+1$,
which is opposite to the case shown in Fig.~\ref{Fig4}. The
experimental results agree well with the theoretical predictions.
From the above experimental results, we can see that our CVV beam
converter can generate any desired CVV beams which is not confined
to the radial and azimuthal cases~\cite{Beresna2011}, and also not
restricted to the topological charge $m=1$. Such a CVV beam
converter enables us to simultaneously manipulate the wavefront and
the polarization state of light.

\section{Conclusions}
In conclusion, we have proposed and realized a simple cylindrical
vector vortex (CVV) beam converter with two cascaded metasurfaces.
Based on spin-orbit interaction, the first metasurface can switch
the topological charges the between positive and negative vortexes
by controlling the handedness of incident polarization. The second
metasurface can manipulate the local polarization orientation of
vector beam. The whole system forms a compact optical system that
provides a simple method to produce CVV beams with any desirable
polarization and vortex phase. The ability to simultaneously tailor
the wavefront and the polarization state of light is expected to
hold potential applications in the fields of optical communication
and mode switching.

\section*{Acknowledgments} This research was partially
supported by the National Natural Science Foundation of China
(Grants Nos. 61025024, 11274106 and 11347120), Hunan Provincial
Natural Science Foundation of China (Grant No. 12JJ7005), and Hubei
Engineering University Research Foundation (Grant No. z2011016).


\begin{thebibliography}{99}

\bibitem{Zhan2009} Q. Zhan, ``Cylindrical vector beams: from mathematical
concepts to applications,'' Adv. Opt. Photon. \textbf{1}, 1-57
(2009).

\bibitem{Youngworth2000} K. S. Youngworth and T. G. Brown, ``Focusing of high numerical
aperture cylindrical vector beams,'' Opt. Express \textbf{7}, 77-87
(2000).

\bibitem{Zhan2002} Q. Zhan and J. R. Leger, ``Focus shaping using cylindrical vector
beams,'' Opt. Express \textbf{10}, 324-331 (2002).

\bibitem{Dorn2003} R. Dorn, S Qubis, and G. Leuchs, ``Sharper focus for a radially
polarized light beam,'' Phys. Rev. Lett. \textbf{91}, 233901,
(2003).

\bibitem{Zhan2004} Q. Zhan, ``Trapping metallic Rayleigh particles with radial
polarization,'' Opt. Express \textbf{12}, 3377-3382 (2004).

\bibitem{Kawauchi2007} H. Kawauchi, K. Yonezawa, Y. Kozawa, and S. Sato, ``Calculation of
optical trapping forces on a dielectric sphere in the ray optics
regime produced by a radially polarized laser beam,'' Opt. Lett.
\textbf{32}, 1839-1841 (2007).

\bibitem{Deng2007} D. Deng and Q. Guo,``Analytical vectorial structure of radially polarized
light beams,'' Opt. Lett. \textbf{32}, 2711-2713 (2007).

\bibitem{Kozawa2010} Y. Kozawa and S. Sato,
``Optical trapping of micrometer-sized dielectric particles by
cylindrical vector beams,'' Opt. Express \textbf{18}, 10828-10833
(2010).

\bibitem{Ye2014} H. Ye, C. Wan, K. Huang, T. Han, J. Teng, Y. S. Ping,
and C. Qiu, ``Creation of vectorial bottle-hollow beam using
radially or azimuthally polarized light,'' Opt. Lett. \textbf{39},
630-633 (2014).


\bibitem{Padgett1999}  M. J. Padgett and J. Courtial, ``Poincar\'{e}-sphere equivalent for light
beams containing orbital angular momentum,'' Opt. Lett. \textbf{24},
430-432 (1999).

\bibitem{Galvez2003} E. J. Galvez, P. R. Crawford, H. I. Sztul, M. J. Pysher, P. J.
Haglin, and R. E. Williams, ``Geometric phase associated with mode
transformations of optical beams bearing orbital angular momentum,''
Phys. Rev. Lett. \textbf{90}, 203901 (2003).

\bibitem{Karimi2010} E. Karimi, S. Slussarenko, B. Piccirillo, L. Marrucci, and E.
Santamato, ``Polarization-controlled evolution of light transverse
modes and associated pancharatnam geometric phase in orbital angular
momentum,'' Phys. Rev. A \textbf{81}, 053813 (2010).

\bibitem{Holleczek2011} A. Holleczek, A. Aiello, C. Gabriel, C. Marquardt, G. Leuchs,
``Classical and quantum properties of cylindrically polarized states
of light,'' Opt. Express \textbf{19}, 9714-9736 (2011).

\bibitem{Milione2011} G. Milione, H. I. Sztul, D. A. Nolan, and R. R.
Alfano, ``Higher-Order Poincar\'{e} Sphere, Stokes Parameters, and
the Angular Momentum of Light,'' Phys. Rev. Lett. \textbf{107},
053601 (2011).

\bibitem{Liu2014} Y. Liu, X. Ling, X. Yi, X. Zhou, H. Luo, and S. Wen
``Realization of polarization evolution on higher-order Poincar\'{e}
sphere with metasurface,'' Appl. Phys. Lett. \textbf{104}, 191110
(2014).

\bibitem{Allen1992} L. Allen, M. W. Beijersbergen, R. J. C. Spreeuw, and J. P.
Woerdman, ``Orbital angular momentum of light and the transformation
of Laguerre-Gaussian laser modes,'' Phys. Rev. A \textbf{45},
8185-8189 (1992).

\bibitem{Yao2011} A. M. Yao and M. J. Padgett, ``Orbital angular momentum: origins,
behavior and applications,''  Adv. Opt. Photon. \textbf{3}, 161-204
(2011).


\bibitem{Zhao2013} Z. Zhao, J. Wang, S. Li, and A. E. Willner,
``Metamaterials-based broadband generation of orbital angular
momentum carrying vector beams,'' Opt. Lett. \textbf{38}, 932-934
(2013).

\bibitem{Cardano2012} F. Cardano, E. Karimi, S. Slussarenko, L. Marrucci, C. de Lisio, and E. Santamato,
``Polarization pattern of vector vortex beams generated by q-plates
with different topological charges,'' Appl. Opt. \textbf{51}, C1-C6
(2012).

\bibitem{Viswanathan2009} N. K. Viswanathan and V. V. G. K. Inavalli,
``Generation of optical vector beams using a two-mode fiber,'' Opt.
Lett. \textbf{34}, 1189-1191 (2009).

\bibitem{Chen2011} H. Chen, J. Hao, B. F. Zhang, J. Xu, J. Ding, and H. T. Wang,
``Generation of vector beam with space-variant distribution of both
polarization and phase,'' Opt. Lett. \textbf{36}, 3179-3181. (2011).

\bibitem{Niv2006} A. Niv, G. Biener, V. Kleiner, and E. Hasman,
``Manipulation of the Pancharatnam phase in vectorial vortices,''
Opt. Express \textbf{14}, 4208-4220 (2006).


\bibitem{Galvez2012} E. J. Galvez, S. Khadka, W. H. Schubert, and S. Nomoto,
``Poincar\'{e}-beam patterns produced by nonseparable superpositions
of Laguerre-Gauss and polarization modes of light,'' Appl. Opt.
\textbf{51}, 2925-2934 (2012).

\bibitem{Litchinitser2012} N. M. Litchinitser, ``Structured light meets structured matter,''
Science \textbf{337}, 1054-1055 (2012).

\bibitem{Kildishev2013} A. V. Kildishev, A. Boltasseva, and V. M. Shalaev,
``Planar photonics with metasurfaces,'' Science \textbf{339},
1232009 (2013).

\bibitem{Beresna2011}  M. Beresna, M. Gecevi\v{c}ius, P. G. Kazansky, and T. Gertus,
``Radially polarized optical vortex converter created by femtosecond
laser nanostructuring of glass,'' Appl. Phys. Lett. \textbf{98},
201101 (2011).


\bibitem{Bomzon2002} Z. Bomzon, G. Biener, V. Kleiner, and E. Hasman,
``Radially and azimuthally polarized beams generated by
space-variant dielectric subwavelength gratings,'' Opt. Lett.
\textbf{27}, 285-287 (2002).

\bibitem{Levy2004} U. Levy, C. Tsai, L. Pang, and Y. Fainman,
``Engineering space-variant inhomogeneous media for polarization
control,'' Opt. Lett. \textbf{29}, 1718-1720 (2004).

\bibitem{Lerman2011} G. M. Lerman and U. Levy, ``Generation of a radially polarized light
beam using space-variant subwavelength gratings at 1064 nm,'' Opt.
Lett. \textbf{33}, 2782-2784 (2008).

\bibitem{Niv2008} A. Niv, Y. Gorodetski, V. Kleiner, and E. Hasman,
``Topological spin-orbit interaction of light in anisotropic
inhomogeneous subwavelength structures,'' Opt. Lett. \textbf{33},
2910-2912 (2008).

\bibitem{Marrucci2006} L. Marrucci, C. Manzo, and D. Paparo,
``Optical Spin-to-Orbital Angular Momentum Conversion in
Inhomogeneous Anisotropic Media,'' Phys. Rev. Lett. \textbf{96},
163905 (2006).

\bibitem{Ling2012} X. Ling, X. Zhou, H. Luo, and S. Wen,
``Steering far-field spin-dependent splitting of light by
inhomogeneous anisotropic media,'' Phys. Rev. A \textbf{86}, 053824
(2012).

\bibitem{Marrucci2011} L. Marrucci, E. Karimi, S. Slussarenko, B Piccirillo, E. Santamato,
E. Nagali, and F. Sciarrino, ``Spin-to-orbital conversion of the
angular momentum of light and its classical and quantum
applications'' J. Opt. \textbf{13}, 064001, (2011).


\bibitem{Born1997} M. Born and E. Wolf, \textit{Principles of Optics} (University Press, Cambridge, 1997).

\end{thebibliography}
\end{document}